\documentstyle[12pt,bezier]{article}

\let\a=\alpha
\let\b=\beta
\let\g=\gamma
\let\d=\delta
\let\e=\epsilon

\let\h=\eta

\let\l=\lambda
\let\m=\mu
\let\n=\nu

\def\p{\vec{p}}
\def\q{\vec{q}}
\let\r=\rho
\let\s=\sigma

\let\o=\omega

\let\S=\Sigma

\let\L=\Lambda
\let\G=\Gamma
\let\D=\Delta

\let\u=\underline

\def\2{{1\over2}} \def\4{{1\over4}} \def\52{{5\over2}}
\def\6{\partial }

\def\({\left(} \def\){\right)} \def\<{\langle } \def\>{
\rangle }
  \def\lb{\left\{} \def\rb{
\right\}}

  \def\CD{{\cal D}}

\def\beg{\begin{equation}}
\def\begar{\begin{eqnarray}}
\def\ee{\end{equation}}
\def\ea{\end{eqnarray}}

\newcommand{\pref}[1]{(\ref{#1})}
\newcommand{\plabel}[1]{\label{#1}}
\newcommand{\prefeq}[1]{Eq.~(\ref{#1} )}

\newcommand{\pcite}[1]{\cite{#1}}
\newcommand{\pbib}[1]{\bibitem{#1}}

\begin{document}

\begin{titlepage}
\hfill TUW-93-13
 \begin{center}
\Large
 {\bf Toponium as Nonabelian Positronium ?}\\[2cm]
\normalsize
 W. Kummer and W. M\"odritsch\\[1.2cm]
 Institut f\"ur Theoretische Physik, TU-Wien\\
 1040 Wien, Wiedner Hauptstra\ss e 8-10/136\\
 FAX: (01143) 222-567760\\
\end{center}
\vspace{1.2cm}
\centerline{Abstract}
The expected large mass $m_t$ of the top quark provides for the
first time a chance to discuss the bound-states of the
corresponding quantum system as the nonabelian generalization of
positronium, using the full power of relativistic quantum field
theoretic methods which are available for weakly bound systems.
Thus our approach differs in principle from the one used in the
vast phenomenological literature on quarkonium potentials. We
emphasize especially the corrections of energy levels which are
of order $\a^4 m_t$ or numerically comparable to that order, and
which have no counterpart in the 'relativistic' corrections of
QED. In contrast to previous computations we give analytic
expressions for all contributions considered in our present
work, hopefully preparing the ground for further similar
calculations.\\

\vspace{\fill}

\_\hrulefill \hspace*{8cm} \\
Vienna, June 1993
\end{titlepage}

\section{Introduction}

Perturbative expansions in the coupling constant in quantum
field theory possess two types of applications, the calculation
of scattering processes and the computation of processes
involving weakly bound systems. Many of the successes of quantum
electrodynamics (QED) are, in fact, related to positronium, i.e.
to the second one of the aforementioned applications. The proper
starting point for any bound-state calculation in quantum field
theory is an integral equation, comprising an infinite sum of
Feynman graphs. The Bethe-Salpeter (BS) equation
\pcite{Bethe} fulfills this task and it is well known that in
the limit of
binding energies of $O(\a^2m)$ the Schr\"odinger equation with
static Coulomb
attraction is obtained. The computation of higher order corrections
to the
Bohr-levels, however, turned out to be far from trivial. It was
recognized, though, relatively late that, at least conceptually,
substantial progress with respect to a systematic treatment
results from a consistent use of a perturbation theory geared to
the original BS equation \pcite{Lep77}. In that manner, at the
same time, nonrelativistic expansions as implied by Hamiltonian
approaches with successive Fouldy-Wouthuysen transformations
\pcite{Fein} are avoided. Within the BS-technique, however, it
is desirable to have an exactly solvable zero order equation
different from the Schr\"odinger equation, otherwise e.g. the
approximation procedure for the wave function lacks sufficient
transparency, especially in higher orders. One of the advantages
of the BS approach to perturbation theory is the freedom to
select a different zero order equation. Of course, in that case,
certain corrections included already at the zero level are to be
properly subtracted out in higher orders. An especially useful
zero order equation has been proposed some time ago by Barbieri
and Remiddi (BR equation \pcite{BR}). Still, one of the most
annoying features of all bound-state calculations remains the
pivotal rule played by the Coulomb gauge. In other gauges, e.g.
already the (in QED vanishing) corrections $O(\a^3m)$ of the
Bohr levels imply to take into account an infinte set of Feynman
graphs \pcite{Love}. Only in very special cases, when certain
subsets of graphs can be shown to represent together a
gauge-independent correction, another more suitable gauge may be
choosen.\\ By contrast to QED the vast literature on bound state
problems in quantum chromodynamics (QCD) adheres to a
description of the quark-antiquark system by the Schr\"odinger
equation with corrections 'motivated' by QCD \pcite{ph}. As long
as a relatively small number of parameters suffices for an
adequate phenomenological description of observed quantum
levels, this approach undoubtably has an ample practical
justification. However, again and again certain deviations from
such phenomenological description are reported \pcite{Franz}.
Thus also for this reason a return to more rigorous QCD
arguments remains as desirable as ever. The standard literature
on this subject almost exclusively is based on nonrelativistic
expansions \pcite{Fein} or on the calculation of purely static
Coulomb forces \pcite{Fisch}. Also very often potentials with
higher order corrections as determined from on-shell quarkonia
scattering are used \pcite{scatt}. In that cases relevant
off-shell effects may be even lost which are typical for higher
order corrections. On the other hand, from the point of view of
relativistic quantum field theory as elaborated in the abelian
case of positronium, a similar, more systematic approach seems
desirable, the more so because the basic techniques are well
developed. In addition, at least in one case, namely the decay
of S-wave quarkonium, the result of a full BS-perturbation
calculation \pcite{KumW}, including the QCD corrections to the
bound state wave function, yields a result very different from
the one which considered only the corrections to the quark
antiquark annihilation alone \pcite{BC}.

In this connection the relatively large size of the running
coupling constant even at high energies represents a well known
problem, together with large coefficients from a perturbative
expansion. For this reason e.g. problems arise in the comparison
of the coupling constant as determined from scattering
experiments within the minimal subtraction scheme
($\overline{MS}$), with the coupling constant to be used in a
consistent weak bound-state approach. The philosophy within our
present work will be that the orders of magnitude, as determined
from $\a_{\overline{MS}}$ will be used for estimates, but that
we shall imply a determination of $\a_s$ by some physical
observable (e.g. energy levels, cf. the remarks after eq.
\pref{pi1er} ) of the quarkonium system itself. In that way
delicate correlations of 'genuine' orders of $\a_s$ from {\it
basically} different types of experiments are avoided.

Of course, the quarkonium system also differs profoundly from
positronium because of the confinement of quarks and gluons.
However, the phenomenological success of the nonrelativistic
quarkonium model can be explained by the fact that the bound
states of heavy quarkonia are deep in the Coulomb funnel and
thus sufficiently far away from the confinement part of the
potential.  Estimates in the early 80-s \pcite{Leut} of
nonperturbative effects, describing the 'tail' of confinement by
a gluon condensate \pcite{Shif} suggested that only with quark
masses well above about 50$GeV$ the importance of such
nonperturbative effects for low Bohr quantum numbers may
decrease sufficiently to make perturbative 'field theoretical'
level corrections competitive and observable.  From high
precision electro-weak experiments of the LEP collaborations,
the mass range of the top quark now seems to be etablished to
lie in the range 100-180$GeV$ \pcite{ALEPH}. Thus for the first
time a nonabelian bound state quarkonium system seems to fullfil
the high mass requirement for a genuine field theoretic
approach. Unfortunately the drawback of this situatuion is that
the weak decay $t \to b+W$ broadens the energy levels
\pcite{Kuehn} for increasing mass $m_t$ so that above $m_t
\approx 140 GeV$ individual levels effectively disappear. Even
for $m_t \approx 100GeV$ this broadening of energy levels
certainly already makes the resolution of different Bohr levels
of $O(m \a^2/4n^2) \approx 1 GeV$ (for $\a \approx 0.2$)
difficult to observe. The first field theoretic correction for
nonabelian QCD already starts at $O(m\a^3)$ \cite{Fisch,Dun},
and the corresponding shift depends on the principal and the
angular momentum quantum number. Although one thus has to fully
acknowledge the problems related to experimental observations of
corrections to $O(\a^4 m)$ for the energy levels of the toponium
system, we have been tempted to trust the ingenuity of
experimentalists to eventually arrive at sufficiently precise
data \pcite{Panch}, together with the hope that $m_t$ does not
lie too far above $100GeV$. In that case, for the first time,
genuine field-theoretic consequences of QCD could be tested at
something like a nonabelian positronium system. Of course, the
well known 'relativistic' corrections, corresponding to the same
type of graphs as in the abelian case are present here as well.
However, already considering only vertex corrections, a
difference to the abelian case at $O(\a^4m)$ was discovered a
long time ago \pcite{Dun}. In addition typical other nonabelian
contributions may appear at this level as well. In contrast to
scattering processes, the determination of the order to which a
certain graph contributes in relativistic bound state
perturbation theory in each case requires a special analysis.
Other terms (from QED and weak interaction), incidentally, may
be of the same numerical order $O(\a^4m)$.  The Coulomb gauge
also entails peculiar additional nonlocal interaction terms in
the effective action, appearing e.g. in the path integral
formulation \pcite{Schw}. We also investigate the effect of
those terms here.

In section 2 we recall some basic facts about BS-perturbation
theory and about the BR equation for nonabelian weakly bound
onium-systems.

Tree graphs leading among others to the well known $O(\a^4m)$
corrections are discussed in sect. 3. As indicated already
above, the one loop vacuum polarization (sect. 4.1) provides a
correction term $O(\a^3m)$ in the nonabelian case. Here we take
the opportunity to point out that effects from some of the
lighter quarks (charm, bottom) in the toponium system must be
treated more carefully than it is usually done by including the
quarks only in the number of flavours of a running coupling
constant.

In sect. 4.2 we revisit the old calculation of Duncan
\pcite{Dun} within our present formalism, avoiding some
approximations made in this early computation which allows us to
make even some statements on the term $O(\a^5m)$.  Finally sect.
5 is devoted to an exploration of possible further corrections
of the same numerical order of magnitude as $O(\a^4m)$. We list
several candidates of relevant QCD graphs and we show in cases
of most simple two loop graphs that such corrections may well be
relevant.  Beside these graphs from QCD, also corrections from
the weak interaction and QED may turn out to contribute to this
order, but the nonlocal Schwinger-Christ-Lee type graphs,
peculiar to the Coulomb gauge, are irrelevant to $O(\a^4m)$ as
shown by an explicit calculation.

\section{BS-Perturbation of the BR Equation}
\plabel{Form}

A correct formulation of QCD in Coulomb gauge entails not only
Faddeev-Popov-ghost terms but also the inclusion of nonlocal
interaction terms \pcite{Schw}. Therefore, the full Lagrangian
reads ($a$=1,...,8 for SU(3)):
\begin{equation} \plabel{Lag}
  {\cal L} = -\frac{1}{4} F_{\m\n}^{a} F^{a \m\n} +
\sum_{j=1}^{f} \bar{\Psi}_j (i \g D -m_j) \Psi_j + B^a (\6_j
A_j^a)-\bar{\h}^a \6^i (\d_{ab} \6_i + g f_{abc} A_i^c) \h^b +
v_1 + v_2
\end{equation}
where the Lagrange multiplier $B^a$ guarantees the Coulomb
gauge, and where
\begar
  D_{\m} &=& \6_{\m} - i g T^a A_{\m}^a, \\ F_{\m \n}^a &=&
\6_{\m} A_{\n}^a -\6_{\n} A_{\m}^a + g f_{abc} A_{\m}^b
A_{\n}^c.
\ea
$v_1$ and $v_2$ are given in \pcite{Schw} and are discussed more
explicitely below.  The above Lagrangian will include all
effects of the strong interaction, but, as we will show, QED and
weak corrections may also give a contribution within the
numerical order of our main interest (O($\a^4$)).

In our notation the BS equation in terms of Feynman amplitudes
for $K$ and $S$ reads as
\beg \plabel{allg}
   \chi_{ij}^{BS}(p;P) = -i
S_{ii'}(\frac{P}{2}+p)S_{j'j}(-\frac{P}{2}+p) \int
\frac{d^4p'}{(2\pi)^4} K_{i'j',i''j''}(P,p,p')
\chi_{i''j''}^{BS}(p';P).
\ee
$\chi$ denotes the BS wave function, $S$ the exact fermion
propagators, and $K$ is the sum of all two fermion irreducible
graphs. Furthermore, we have introduced relative momenta $p$ and
$p'$, a total momentum $P=p_1-p_2$, and we choose a frame where
$P=(P_0,\vec{0})$.\\

The notation can be read off the pictorial representation in
fig.1.  $i,j$ are collective indices for spin (noted $\s,\r$)
and colour (noted $\a,\b$).  It is well known that the dominant
part in $K$ for weak binding ($\a \to 0$) is the one-Coulomb
gluon exchange which results in an ordinary Schr\"odinger
equation with static Coulomb potential. This result is even
independent of the chosen gauge in the ladder approximation by a
simple scaling argument $p_0 \approx O(m\a^2), |\p| \approx
O(m\a), P_0 \approx 2m-O(m\a^2)$ \pcite{Kum}:
\beg \plabel{Kc}
  K \to K_c := -\frac{4 \pi \a}{(\p-\p\,')^2} \g^0_{\s \s'}
\g^0_{\r' \r}
\ee
In equation \pref{Kc} we have already used the fact that only
colour singlet states can form bound states because the Coulomb
potential is repulsive for colour octets. The colour trace will
always be understood to be already done, leading to the
definition
\beg
 \a \equiv \frac{4}{3} \frac{g^2}{4\pi} = \frac{4}{3}
\a_s.\plabel{alpha}
\ee
in terms of the usual strong coupling constant $\a_s$.  Because
the above mentioned nonrelativistic limit of the BR equation
contains the projection operators $\l^\pm$, defined below, it is
awkward to calculate the so-called relativistic corrections in a
straightforward way within the framework of BS perturbation
theory, starting from \pref{allg} with \pref{Kc}.  Therefore, we
use the BR equation \pcite{BR} instead of the Schr\"odinger
equation. It is obtained from \pref{allg} by the substitutions
\begar
  S &\to& \frac{1}{\g p -m}, \nonumber \\ K &\to& K_{BR} \equiv
[\g_0 \L^+ \l^+ \L^+]_{\s',\s''} [\L^- \l^- \L^- \g_0
]_{\r',\r''} \tilde{K},  \plabel{BRK}
\ea
with the projection operators
\begar
 \L^{\pm}(\p) &\equiv& \frac{E_p + \vec{\a} \p \pm \b m}{2E_p},
\qquad E_p\equiv \sqrt{\p\,^2+m^2}, \plabel{L} \\ \l^{\pm}
&\equiv& \frac{1}{2}(1 \pm \g_0) \nonumber
\ea
and
\beg  \plabel{Kschl}
 \tilde{K} \equiv - \frac{4\pi \a}{(\p-\p\,')^2} m
\frac{2E_p}{E_p+m} \frac{2E_{p'}}{E_{p'}+m}
\frac{2}{\sqrt{P_0+2E_p}} \frac{2}{\sqrt{P_0+2E_{p'}}}.
\ee
$m$ denotes the mass of the (heavy) quark.  The solutions of the
equations obtained in this manner are exact.  The (colour
singlet, normalized \pcite{Luri}) eigenfunctions read ($ \o :=
E_p - \frac{P_0}{2}$) \newpage
\begar
 \chi(p) &=&
-i\frac{\L^+\l^+\G\l^-\L^-}{(p_0-\o+i\e)(p_0+\o-i\e)} \frac{E_p
\sqrt{2E_p+P_0} (P_0-2E_p)}{\sqrt{P_0}(E_p+m)} \phi(\p)
\nonumber \\ \plabel{xBR} \\ \bar{\chi}(p) &=&
i\frac{\g_0(\L^+\l^+\G\l^-\L^-)^*\g_0}{(p_0-\o+i\e)(p_0+\o-i\e)}
\frac{E_p \sqrt{2E_p+P_0} (P_0-2E_p)}{\sqrt{P_0}(E_p+m)}
\phi^*(\p) \nonumber \\ \nonumber
\ea
and belong to the spectrum of bound states
\beg \plabel{P0}
 P_0 =M_n^0 = 2m \sqrt{1-\frac{\a^2}{4n^2}} \approx 2m - m
\frac{\a^2}{4n^2} - m \frac{\a^4}{64n^4} +O(\a^6).
\ee
In eqs. \pref{xBR} $\G$ is a constant $4\times4$ matrix which
represents the spin state of the particle-antiparticle system:
\beg
      \G = \left\{ \begin{array}{l@{\quad:\quad}l} \g_5 \l^- &
\mbox{S=0} \\ \vec{a}_m \vec{\g} \l^- &  \mbox{S=1}.
\end{array}  \right.
\ee
$\phi$ is simply the normalized solution of the Schr\"odinger
equation in momentum space, depending on the usual quantum
numbers $(n,l,m)$ \pcite{Wf}, $a_{\pm 1}, a_0$ describe the
triplet states.  In the following it will often be sufficient to
use the nonrelativistic approximations of eqs. \pref{xBR}
\begar
  \chi(p)^{nr} &=& \frac{\sqrt{2} i \o}{p_0^2-\o^2+i\e} \phi(\p)
\G \plabel{xapp}, \\
\bar{\chi}(p)^{nr} &=& \frac{\sqrt{2} i \o}{p_0^2-\o^2+i\e}
\phi^*(\p) (-\g_0 \G^* \g_0). \plabel{xqapp}
\ea

Perturbation theory for the BS equation starts from an exactly
solvable equation, in our case the BR equation for the Green
function $G_0$ of the scattering of two fermions \pcite{BR2}
\beg
 iG_0 = -D_0 + D_0 K_0 G_0.
\ee
$D_0$ is the product of two zero order propagators, $K_0$ the
corresponding kernel.  The exact Green function may be
represented as
\beg \plabel{Reihe}
 G =\sum_{l} \chi_{nl}^{BS} \frac{1}{P_0 - M_n}
\bar{\chi}_{nl}^{BS} + G_{reg}=G_0 \sum_{\n =0}^{\infty} (H
G_0)^{\n} ,
\ee
where the corrections are contained in the insertions $H$.
Bound state poles $M_n$ contribute, of course, only for
$P_0<2m$.  It is easy to show how $H$ can be expressed by the
full kernel $K$ and the full propagators $D$:
\beg \plabel{H}
 H = -K + K_0 +iD^{-1}-iD_0^{-1}.
\ee
Since the corrections to the external propagators contribute
only to $O(\a^5)$ \pcite{Male}, the perturbation kernel is
essentially the negative difference of the exact BS-kernel and
of the zero order approximation.

Expanding both sides of equation \pref{Reihe} in powers of
$P_0-M_n^0$, with $M_n^0$ from \pref{P0} the mass shift is
obtained \pcite{Lep77}:
\beg \plabel{dM}
 \D M = \< h_0 \> (1+\< h_1 \> ) + \< h_0 g_1 h_0 \> + O(h^3) .
\ee
Here the BS-expectation values are defined as e.g.
\begar
 \<h\> &\equiv& \int \frac{d^4p}{(2\pi)^4} \int
\frac{d^4p'}{(2\pi)^4} \bar{\chi}_{ij}(p) h_{ii'jj'}(p,p')
\chi_{i'j'}(p'), \plabel{erww}
\ea
We emphasize the four-dimensional p-integrations which
correspond to the generic case, rather than the usual three
dimensional ones in a completely nonrelativistic expansion.

In \pref{dM} $h_i$ and $g_i$ represent the expansion
coefficients of $H$ and $G_0$, respectively, i.e.
\begar
 H&=& \sum_{n=0}^{\infty} h_n (P_0-M_n^0)^n \\ G_0&=&
\sum_{n=0}^{\infty} g_n (P_0-M_n^0)^{n-1}
\ea

\section{QCD Tree Diagrams}
\plabel{tree}

The contributions stemming from the tree diagrams 2.a to 2.c are
well-known from the abelian case.  Fig. 2.a is peculiar for the
use of a different zero order equation than the Schr\"odinger
equation. It contains the difference between the exact one
Coulomb-gluon exchange and the BR-Kernel \pref{BRK}. The
exchange of one transverse gluon is represented by graph 2.b,
and fig. 2.c shows the annihilation graph.  The latter does not
contribute in our nonabelian case.

For the sake of completeness and in order to illustrate the
present formalism, we exhibit first the results for the tree
graphs as well.  The perturbation kernel for the Coulomb-gluon
exchange
\beg
  -iH_c := -ig^2 T^a_{\a \a'} T^a_{\b'\b} \g^0_{\s \s'}
\g^0_{\r' \r} \frac{1}{(\p-\p\,')^2},
\ee
is needed for the calculation of the energy shift induced by
fig. 2.a using Eqs. \pref{dM}, \pref{H}, \pref{erww} and
\pref{BRK} to \pref{Kschl}. For the spin-singlet we have:
\begar
 \D M_{1.a} &=& \<H_c+K_{BR}\> = \nonumber \\ &=&
\frac{\a^2}{16n^2}\<K_c\> + \frac{\pi \a}{P_0 m} \int
\frac{d^3p}{(2\pi)^3} \int \frac{d^3p'}{(2\pi)^3} [\phi^* \phi -
2 \phi^* \frac{\p \p\,'}{(\p-\p\,')^2} \phi] + O(\a^6) =
\nonumber \\ &=& m \a^4( \frac{\d_{l0}}{8n^3} + \frac{1}{16n^4}
- \frac{1}{16 n^3 (l+1/2)}) + O(\a^6) . \plabel{dMc}
\ea

The contribution from the transverse gluon (fig. 1.b)
\beg
 -i H_{1.b} = i 4\pi \a \g^j_{\s\s'} \g^k_{\r'\r} \frac{1}{q^2}
(g_{ik}+\frac{q_j q_k}{\vec{q}\,^2}), \plabel{H1b}
\ee
with
\beg
      q\equiv p'-p,
\ee
gives rise to a spin singlet-triplet (magnetic hyperfine)
splitting.  Because of the $\g^j$ matrices, the $\l^{\pm}$
projectors from both wave functions annihilate \pref{H1b}.  This
means that two factors $\p \vec{\g}$, contained in $\L^\pm$, are
needed for a nonzero result which in turn gives rise to two
extra orders of $\a$. By this mechanism we arrive at the well
known  contribution $O(\a^4)$ from this graph.  For the
spin-singlet the mass shift reads
\begar
\D M_{1.b,S=0}&=& \frac{2\pi \a}{m^2} [-|\Psi(0)|^2 + 2 \int
\frac{d^3p}{(2\pi)^3} \int \frac{d^3p' }{(2\pi)^3}  \phi^*(\p\,')
(\frac{(\p\q)(\p\,'\q)}{\q\,^4}- \frac{\p\p\,' }{\q\,^2}) \phi(\p)
 ]+ O(\a^6) = \nonumber \\
 &=& m\a^4(\frac{1}{8 n^4}
 -\frac{\d_{l0}}{8 n^3} -\frac{3}{16 n^3 (l+\2)})+ O(\a^6).
 \plabel{dMT}
\ea
The evaluation of the singlet-triplet splitting requires some
awkward Dirac-algebra, but the final result may be brought in a
quite transparent form (where one recognizes this expression as
the well known spin-spin and spin-orbit interaction, adapted to
the present problem, cf. e.g. \pcite{Land})
\begar
 \D M_{ortho,m} - \D M_{para} &=& \frac{2 \pi \a}{m^2} \int d^3p
\int d^3p' \phi^*(\p\,')(1+\frac{|\q \vec{a}_m|^2}{\q\,^2}- 3
\frac{(\p\,' \times \p)(\vec{a}^*_m \times \vec{a}_m)} {\q\,^2}
) \phi(\p)= \nonumber\\ &=&    \frac{2 \pi \a}{m^2} \<
\frac{4}{3} \d(\vec{r}) + \frac{1}{4\pi}
\frac{\vec{r}\,^2-3|\vec{r} \vec{a}_m|^2}{|\vec{r}|^5} -
\frac{3i(\vec{r} \times \p)(\vec{a}^*_m \times \vec{a}_m)}{4\pi
r\,^3} \>.\plabel{op}
\ea
For our purpose it is sufficient to calculate the
singlet-triplet splitting for the ground state given by the term
proportional to $\d(\vec{r})$:
\beg \plabel{ortp}
 \D M_{ortho} - \D M_{para} = \frac{8 \pi \a}{3 m^2}
|\Psi(0)|^2=m \a^4 \frac{\d_{l0}}{3n^3}.
\ee
It should be mentioned that this result differs from the
positronium case, because the graph 2.c gives no contribution
here. Moreover, we want to emphasize again that according to our
'purist' field theoretic point of view, integrals like \pref{op}
and \pref{ortp}, of course, are {\it not} to be evaluated
between phenomenological wave-functions.

\section{One Loop Corrections}

\subsection{One Loop Vacuum Polarization}

In the case of positronium no massless particles can contribute
to vacuum polarization and so this effect is only of order of
magnitude $\a^5$.  In contrast to this, QCD contains massless
gluons and light quark flavours which may contribute
significantly to the spectrum.

Using standard techniques for evaluating the corresponding
integrals one obtains for the loop in fig. 3.a ($N$=3 for QCD):
\begar
 \pi^{ab}_{3.a} &\equiv& 4 g^2 N \d^{ab} \int
\frac{d^Dr}{(2\pi)^D} \frac{\q\,^2 - \frac{(\q
\vec{r})^2}{\vec{r}^2}}{r^2 (\q-\vec{r})^2 } \nonumber \\ &=&
-\frac{i g^2 N \d^{ab}}{3 \pi^2} \q\,^2 [ \CD -\ln\q\,^2
+\frac{7}{3} - 2\ln2 -  \plabel{pi2er}\\ & & - \e(\frac{7}{3} -
2\ln2 -\g +\ln 4\pi) \ln \q\,^2 + \frac{\e}{2} \ln^2\q\,^2 + \e
\cdot const + O(\e^2)] \nonumber
\ea
with
\begar
     \CD &=& \frac{1}{\e}-\g+\ln 4\pi, \qquad \e = \frac{4-D}{2}
\plabel{Div}
\ea
We have written down this result correct to orders $\e\cdot
f(q)$ in order to make it applicable in the two loop
calculation.  The graph 3.b is more difficult to calculate,
because it contains $q_0$ terms. These terms can be completely
avoided if we carry out the $p_0$ integrations first (cf.
\pref{erww} and \pref{dM}).  The result can be expanded in
powers of $\p$ and $\a$ in order to show that the effect of
$q_0$ is of $O(\a^5)$.  With this simplification graph 3.b is
exactly calculable:
\beg \plabel{pi1er}
 \pi^{ab}_{3.b} = \frac{i g^2 N \d^{ab}}{96 \pi^2}[ 10 \q\,^2
(\CD -\ln\q\,^2) +16(7 - 8\ln2) \q\,^2 ] +O(\e)
\ee
Our renormalization prescription consists of a subtraction at
the point $q=(0,\vec{\m})$, where $\m$ has to be of the order
$\a m$ to avoid large logarithmic contributions from higher
orders. This seems to be the natural renormalization
prescription for bound state problems, because also in the BS
expectation values \pref{erww}, the Bohr momentum $\a m$
together with $p_0 \approx O(\a^2 m)$ provides the dominant
parts of the integrals.

After renormalization, the contribution from the gluonic vacuum
polarization (with the colour trace already done) reads
\pcite{Dun}
\begar \plabel{Kg}
-iH_g &=& -i \g_0 \otimes \g_0 \tilde{H_g}\\
\tilde{H_g} &=&-\frac{ 11 \a^2 N}{4 \q\,^2} \ln \frac{\q\,^2}
{\vec{\m}^2}.
            \nonumber
\ea
The expectation value of this expression can be obtained by
performing the Fourier transformation into coordinate space,
where the integrations can be done analytically (see Appendix
A).  Our surprisingly simple result is
\beg \plabel{dMg}
 \D M_g = \<H_g\> = - m \a^3 \frac{11N}{16 \pi n^2 } [
\Psi_1(n+l+1)+\g + \ln \frac{\m n}{\a m}] + O(\a^5)
\ee
where $\Psi_n$ is the n-th logarithmic derivation of the gamma
function and $\g$ denotes Euler's constant. The closed form of
Eq. \pref{dMg} was not obtained in previous calculations.

Now we turn to the contribution from the fermion loops fig. 3.c.
In the literature the lighter quarks are usually taken as
massless (and 'absorbed' in the number of flavours appearing in
the running coupling constant) or even ignored \pcite{Dun}
\pcite{Fein}, but we will show that they do contribute within
the order of interest and, furthermore, the explicit dependence
on the masses of the lighter quarks is important.  As pointed
out already in the introduction, this is due to the fact that
the top quark is expected to lie above 100GeV \pcite{Barb} and
therefore the bottom and charm quark can neither be taken as
relatively massless nor as relatively super-heavy as compared to
the natural mass scale $\a m$.

The finite part of the self energy in fig. 3.c is a well known
quantity
\pcite{Land} for arbitary masses $m_j$ of the quark:
\beg
 \Pi_F = -\frac{i g^2 \d^{ab}}{4\pi^2} \q\,^2 I(\q\,^2,m_j^2).
\ee
We approximate in the exact solution
\begar
 I(\q\,^2,m_j^2) &\equiv& \int_0^1 dx x(1-x) \ln
[x(1-x)\q\,^2+m_j^2] \nonumber \\ &=& \frac{1}{6} \ln m_j^2
-\frac{5}{18} + f(\r), \\ f(\r) &=& \frac{2 \r}{3} + \frac{1}{6}
(1-2\r) \sqrt{1+4\r} \ln \frac{\sqrt{4 \r +1}+1}{\sqrt{4 \r
+1}-1} \plabel{f},\\ \r &:=& \frac{m_j^2}{\q\,^2},
\ea
for later convenience $f(\r)$ by
\beg \plabel{fnaeh}
 f(\r) \approx \frac{1}{6} \ln (\frac{1}{\r} + e^{\frac{5}{3}}).
\ee
This agrees with the original $f(\r)$ better than 1\% within the
whole integration region.

It seems instructive to transform into coordinate space in order
to obtain the potential, effectively produced by this fermionic
vacuum polarization:
\begar
 \D M_F &=& \< H_F \>, \nonumber\\ H_F(r) &=& -\frac{\a^2}{4\pi
r} [\mbox{Ei}(-r m_j e^{\frac{5}{6}}) -\frac{5}{6} + \2 \ln
(\frac{\m^2}{m_j^2} + e^{\frac{5}{3}})].  \plabel{HF}
\ea
The mass shift can be obtained from \pref{HF} in closed form
using the integral formula \pcite{Grad}
\begar \plabel{Eiint}
 \int_0^x e^{-\b x} \mbox{Ei}(-\a x) dx &=& -\frac{1}{\b} [
e^{-\b x} \mbox{Ei}(-\a x)+ \ln (1+\frac{\b}{\a}) - \\ & & -
\mbox{Ei}(-(\a+\b)x) ]. \nonumber
\ea
Thus a useful expression for the energy shift induced by
fermionic vacuum polarization with {\it arbitrary} masses $m_j$
reads
\begar
 \D M_F^j &=& -\frac{m\a^3}{8\pi n^2 } \lb \sum_{k=0}^{2n-2l-2}
b_{nl}^k \left[ \left( -\frac{d}{d\b} \right)^{2l+1+k}
[-\frac{1}{\b} \ln (1+\b \frac{\a m}{n m_j e^{\frac{5}{6}}})]
\right]_{\b \to 1} - \right. \nonumber \\ & & \qquad \left.
-\frac{5}{6} + \2 \ln (\frac{\m^2}{m_j^2} + e^{\frac{5}{3}})
\rb,  \plabel{KF}
\ea
with
\beg
  b_{nl}^k := \frac{(n-l-1)!}{k![(n+l)!]^3} \left(\frac{d}{d\r}
\right)^k [L_{n-l-1}^{2l+1}(\r)]^2 \Big|_{\r=0} .
\ee
For states up to $n=3$ we write this result more explicitly as
\beg
 \D M_{F,nl}^j = \frac{m\a^3}{8\pi n^2 } \lb A_{nl}(\frac{n
m_j}{\a m})+ \ln \frac{(\frac{\m^2}{m_j^2}+
e^{\frac{5}{3}})^{\2}}{e^{\frac{5}{6}} + \frac{\a m}{n m_j} }
\rb \plabel{KFex}
\ee
with $A_{nl}$ from Tab.2,\\[.5cm]
\begin{minipage}{16cm}
\centerline{\begin{tabular}{c|c|c}
 $n$ & $l$ & $A_{ln}(\frac{n m_j}{\a m})$ \\ \hline 1 & 0 & $a$
\\ 2 & 0 & $a^3-\2 a^2 + a$\\ 2 & 1 & $\frac{1}{3}a^3+\2
a^2+a$\\ 3 & 0 & $2 a^5-\frac{7}{2}a^4+\frac{10}{3}a^3-a^2+a$\\
3 & 1 & $a^5 - \frac{3}{4} a^4 + \frac{1}{3} a^3 + \2 a^2 + a$\\
3 & 2 & $\frac{1}{5}a^5 +\frac{1}{4}a^4 +\frac{1}{3}a^3+\2 a^2 +
a$
\end{tabular}}
\vspace{.3cm}
\centerline{Tab. 2}
\vspace{.5cm}
\end{minipage}
using the shorthand
\beg
 a^{-1}:= 1+\frac{ n m_j e^{\frac{5}{6}} }{\a m}
\ee
Only for $m_j>>\a m$ this gives an $O(\a^5)$ Uehling term,
modified by off-shell subtraction, but for $m_j \to 0$ it
becomes an $O(\a^3)$ contribution, which means that eq.
\pref{KFex} interpolates numerically in a range of two orders in
$\a$. Therefore \pref{KFex} must be definitely taken into
account for quarks with $m_j \approx \a m$ at $O(\a^4)$.

\setcounter{equation}{0}

\section{Vertex Corrections}

The one loop corrections to the vertex together with self-energy
insertions into the fermion lines (fig. 3.d) in the abelian case
(positronium), are known to provide corrections only of
$O(\a^5)$. The reason for this is the "classical" Ward identity
which continues to relate those contributions in such a way that
the sum of these terms vanishes at $|\q| \to 0 $. This Ward
identity happens to continue to hold even in the Coulomb gauge
and even in the nonabelian case
\pcite{Fein}, but only for the vertex
corrections referring to the Coulomb component of the gauge
field. However, the presence of the gluon splitting graphs 3.e
and 3.f produces a contribution already to $O(\a^4)$
\pcite{Dun}. A simple dispersion theoretic argument allows to
understand this difference: In the abelian case the first graph
3.d in the variable $|\q|$ for $q_0=0$ has a cut for
Re$|\q|<2m$. Thus corrections in $\q$, e.g. in the electron form
factor $F_1$, for symmetry reasons must be of order $\q\,^2$,
because $F_1(\q\,^2)$ is regular at $\q \to 0$. This is no
longer the case with the mass zero intermediate state allowed in
3.e and 3.f. The first - and to our knowledge only- computation
of the nonabelian vertex corrections in the sense of our present
approach was performed in ref. \pcite{Dun}. This work contains
certain approximations which we wanted to avoid in order to
prepare the ground for future calculations even at the level
$O(\a^5)$. We thus make a systematic expansion and solve the
remaining integrals analytically which contain contributions of
the order of interest.  The vertex correction of fig. 3.e after
performing the colour trace becomes
\begar
 -iH_{3.e} &=& \frac{36 \pi^2 \a^2 q_i}{\q\,^2} \int
\frac{d^4r}{(2\pi)^4}  \frac{1}{r^2 (\vec{r}-\q)^2}
(-\d_{ki}+\frac{r_k r_i}{\vec{r}^2}) \times \nonumber \\ & &
\times [\g^k \frac{1}{\g p + \g r -m} \g^0 - \g^0 \frac{1}{\g p'
-\g r -m} \g^k] \nonumber \\ &=& \frac{36 \pi^2 \a^2
q_i}{\q\,^2} ( \g^k v_{ki}(1,p)\g^0-\g^0 v_{ki}(-1,p') \g^k)
\plabel{Vk}
\ea
where ( $\e = \pm 1$)
\begar
  v_{ki}(\e,p) &:=& \int \frac{d^4r}{(2\pi)^4}  \frac{1}{r^2
(\vec{r}-\q)^2} (-\d_{ki}+\frac{r_k r_i}{\vec{r}^2}) \frac{\g p
+ \e \g r +m}{(p+\e r)^2 -m^2} \plabel{vki}
\ea
After the $r_0$ integration it proves useful to proceed with the
$p_0$ integrations (cf. \prefeq{erww} ), where in contrast to
ref. \pcite{Dun}, who approximates already at this point, we
took into account also the pole arising from the denominator of
Eq. \pref{vki}. This results in
\begar
 v_{ki}(\e,p) &=& -i \int \frac{d^3r}{(2\pi)^3}
\frac{-\d_{ki}+\frac{r_k r_i}{\vec{r}^2}}{(\vec{r}-\q)^2}
F(\p,\vec{r}) \\ F(\p,\vec{r}) &=& \frac{[(r+E_{p+\e r})(m
-\vec{\g}\p - \e \vec{\g} \vec{r}) +\g_0 E_{p+\e r}
\frac{P_0}{2}](r+E_{p+\e r}+\o)- \g_0 E_{p+\e r} \frac{P_0}{2}
\o} {2 r E_{p+\e r} (r+ E_{p+\e r})[ \frac{P_0^2}{4} - (r+
E_{\p+\e\vec{r}} +\o)^2]}. \nonumber \\
\ea
The calculation of the $\g$ trace to lowest order requires the
inclusion of the $\vec{\g}\p$-terms in the wave functions
\pref{xBR} stemming from the projection operators \pref{L}.
After performing the $\g$ trace we expand in terms of $\p$ and
$m\a^2$ which enables us to combine both terms in eq. \pref{Vk}:
\beg
 H_{3.e} = \frac{36 \pi^2 \a^2}{\q\,^2} \int
\frac{d^3r}{(2\pi)^3} \frac{\q\,^2-\frac{(\q
\vec{r})^2}{r^2}}{(\vec{r}-\q)^2} \frac{1}{2 r^2 E_r} +
\mbox{higher orders}   \plabel{Vnaeh}
\ee
Now Eq. \pref{Vnaeh} may be evaluated exactly in terms of
dilogarithms and the result has a cut for $|\q|<0$, but no pole
at $|\q|=0$. It can be formally expanded for $|\q|>0$ to
$O(|\q|)$:
\beg
 H_{3.e} = \frac{9 \a^2}{|\q| m } (\frac{\pi^2}{8}-\frac{2
|\q|}{3}+ O(\q\,^2) ). \plabel{Va}
\ee
The first term  in this expansion has been obtained in
\pcite{Dun}, the second one is the expected contribution to
$O(\a^5)$.  The BS - expectation value of \pref{Va} becomes to
$O(\a^4)$:
\beg
 \D M_{3.e} = \frac{9 m \a^4}{36 n^2} \frac{1}{n(l+\2)}.
\plabel{nakorr}
\ee

It remains to calculate the vertex corrections with two
transverse gluons , graph 3.f. At first sight it seems that this
graph would give a contribution to order $\a^3$ because no $\p
\vec{\g}$ terms are needed from the wave functions. This, as we
will see, is not true because the leading (constant) term
vanishes as a consequence of renormalization and accordingly
graph 3.f has been estimated to be of order $O(\a^5)$ in ref.
\pcite{Dun}. Here we use an approach which explicitely provides
at least part of the exact result of this contribution.  The
vertex part of the graph 3.f reads:
\begar \plabel{Vb}
V_{3.f} &=& -\frac{N g^3 T^b}{4} \int \frac{d^Dr}{(2\pi)^D}
\frac{(q_0-2r_0) \g^i (\g(p-r)+m) \g^k} {[(p-r)^2-m^2](r-q)^2
r^2} \times \\ & & \qquad \qquad \times \left(
\d_{ik}-\frac{(q_i-r_i)(q_k-r_k)}{(\q-\vec{r})^2} -\frac{r_i
r_k}{\vec{r}^2} + \frac{(q_i-r_i) \vec{r}(\q-\vec{r}) r_k}
{\vec{r}^2 (\q-\vec{r})^2} \right) \nonumber
\ea
With the gamma-trace to relative $O(\a^2)$ and using Feynman
parametrization, the effective vertex part becomes
\begar
 V_{3.f}^{eff} &=& -\frac{i g^3 N T^c}{16} \int
\frac{d^{D-1}k}{(2\pi)^{D-1}} \int_0^1 dx \int_0^{1-x} dy
(L^{-\frac{3}{2}}-3x^2m^2L^{-\frac{5}{2}})(1+\frac{[\vec{k}
(\q-\vec{k})]^2}
{\vec{k}^2(\q-\vec{k})^2}), \plabel{vbna} \nonumber \\
\ea
where
\begar
              L &=& (y
q_0-xp_0)^2+\vec{k}^2+2x\p\vec{k}-2y\q\vec{k}-x(p^2-m^2)-y
\q\,^2.
\ea
For our estimate it is sufficient to consider the part stemming
from the constant term in the second bracket in \prefeq{vbna}.
The $\vec{k}$ integration can be done easily leaving us with a
finite part
\begar
 V_{3.f,1}^{eff} &=&\frac{i g^3 N T^c}{32 \pi^2} \int_0^1 dx
\int_0^{1-x} dy (\ln M+\frac{2 x^2 m^2}{M})
\ea
where
\begar
         M &=& x^2 m^2 - w,\\ w
&=&y(1-y)q^2+x(1-x)(p^2-m^2)+2xypq.
\ea
This expression cannot be expanded in terms of $w$ because this
would yield a spurious linear divergence in the next order from
the $q^2$ term which would indicate an equally spurious
$O(\a^4)$ contribution. Therefore, we expand in terms of
$(w+y(1-y)\q\,^2)$ and solve the leading part analytically in
terms of dilogarithms. Expanding the result in terms of
$\q\,^2$, one has
\begar
     V^{eff}_{3.e} &=&\frac{i g^3 N T^c}{32
\pi^2}\{2-\frac{1}{4}[1+\ln(
\frac{\q\,^2}{m^2})]\frac{\q\,^2}{m^2} \}+ \mbox{other
$O(\a^2)$} \plabel{Veff}
\ea
Since \pref{Veff} does not contains a term $\propto |\q|$, the
vertex correction due to two transverse gluons does not
contribute to $O(\a^4)$.

\subsection{Other QCD graphs}

To one loop order also the box graph 3.g occurs in the
correction to the BR kernel. It possesses an exact counterpart
in QED and is known to contribute only to $O(\a^5)$
\pcite{Fult}.  Furthermore we have also investigated the
two-loop vertex-correction depicted in fig. 3.h. Of course this
correction is but one of several two loop vertex corrections.
The renormalization must take into account the whole set of this
graphs. Nevertheless, it seems that this graph, after proper
renormalization, yields an $O(\a^5)$ contribution.

\setcounter{equation}{0}

\section{Other Corrections}

\subsection{Two Loop Vacuum Polarization}

As pointed out already in subsection 4.1, the usual
renormalization group arguments relying on massless quarks in
the running coupling constant do not consistently include the
effect of 'realistic' quark masses in the toponium system, when
a systematic BS perturbation is attempted. However, in the one
loop case finite quark masses gave terms of numerical order
$O(\a^4)$, therefore the same may be expected here, leading to
corrections of $O(\a^5)$.  On the other hand, in a full
calculation of effects of $O(\a^4)$ two loops with gluons cannot
be neglected. Altough it is enough to consider the two loop
vacuum polarization for Coulomb gluons only, the computation of
all those graphs is beyond the scope of our present paper. We
just want to indicate how already the graphs 4.a-4.c yield
contributions of this order.  The result of the self energy part
from 4.a to 4.c are given in tab.3.  Let us start with graph
4.a.  Performing the zero component integrations of momenta
results in (including a symmetry factor 1/2)
\begar
 -iH_{4.a} &=& -i \g_0 \otimes \g_0 \frac{g^6 N^2}{2 \q^4}
(\Pi^{(1)}+\Pi^{(2)}) \ea with
\begar
\Pi^{(1)} &=& \int \frac{d^{D-1}k}{(2\pi)^{D-1}} \int
\frac{d^{D-1}p}{(2\pi)^{D-1}}
              \frac{1}{\p\,^2 |\vec{k}| |\q-\vec{k}-\p|},\\
\Pi^{(2)} &=& \int \frac{d^{D-1}k}{(2\pi)^{D-1}} \int
\frac{d^{D-1}p}{(2\pi)^{D-1}}
              \frac{[(\q-\vec{k}-\p)\vec{k}]^2}{\p\,^2
|\vec{k}|^3 |\q-\vec{k}-\p|^3}.
\ea
By using dimensional regularization, Feynman-parametrization and
usual integration formulas \pcite{Muta} we arrive at
\begar
 \Pi^{(1)} &=& \frac{\G(-\2+\e) \G^2(1-\e)}{\G^2(\2)
(4\pi)^{\frac{3}{2}-\e} \G(2-2\e)} \int
\frac{d^{D-1}p}{(2\pi)^{D-1}} \frac{1}{\p\,^2
[(\q-\p)^2]^{-\2+\e}}= \plabel{Pi1} \\ &=& \q\,^2
\frac{(4\pi)^{2\e} (\q\,^2)^{-2\e}}{\G^2(\2) (4\pi)^3}
\frac{\G^2(1-\e) \G(-1+2\e) \G(\2-\e)}{\G(\frac{5}{2}-3\e)}
\nonumber
\ea
and
\begar
 \Pi^{(2)} &=& \frac{3 \G(\frac{3}{2}-\e) \G^2(1-\e)
\G(2\e)}{4(4\pi)^{3-2\e}(\frac{3}{2}-2\e)(-1+2\e) \G(\52) \G(\2)
\G(\52-3\e)} (\q\,^2)^{1-2\e}+ \nonumber\\ & & + \frac{3
\G(\2-\e) \G(2-\e) (\q\,^2)^{1-2\e}}{2 (4\pi)^{3-2\e}
\G(\frac{3}{2}) \G(\52)} \times \plabel{Pi2} \\ & & \lb
\frac{\G(2\e) \G(\frac{7}{2} - \e) \G(2-\e)}{(-1+2\e)
\G(\frac{3}{2}-\e) \G(\frac{9}{2} -3\e)} + \frac{\G(2\e)}{3-4\e}
\left[ \frac{\G(1-\e)}{\G(\52-3\e)} -
\frac{4\G(2-\e)}{\G(\frac{7}{2}-3\e)}+\frac{4\G(3-\e)}{\G(
\frac{9}{2}-3\e)}
\right] \right. \nonumber \\ & & \left.\qquad \qquad - \frac{2
\G(2\e) \G(\52-\e) \G(2-\e)}{\G(\frac{3}{2}-\e)
\G(\frac{9}{2}-3\e)}+ \frac{\G(2-\e)
\G(1+2\e)}{\G(\frac{9}{2}-3\e)} \rb. \nonumber
\ea
{}From eqs. \pref{Pi1} and \pref{Pi2} the finite, renormalized
contribution to the perturbation kernel may be extracted as
\beg
 H_{4.a} = \g_0 \otimes \g_0 \frac{81 \a^3}{8 \pi \q\,^2} \ln
\frac{\q\,^2}{\vec{\m}^2}.  \plabel{H3a}
\ee
Eq. \pref{H3a} differs from eq. \pref{Kg} by a simple factor
proportional to $\a$ and therefore results in the mass shift
\beg  \plabel{dM4}
 \D M_4 = \frac{81 m \a^4}{64 \pi^2 n^2 }(\Psi_1(n+l+1)+\g + \ln
\frac{\m n}{\a m}).
\ee
The contribution from fig. 4.b is similar. After performing the
integrations over the zero components we have
\begar
 \Pi_{4.b} &=& -\frac{i g^4 3 N^2 \d_{ab}}{2} \int
\frac{d^{D-1}k}{(2\pi)^{D-1}} \int \frac{d^{D-1}p}{(2\pi)^{D-1}}
\frac{1}{(\q-\vec{k})^2 |\p| |\vec{k}| (\q-\p)^2} \times  \\ & &
\times \left[\q\,^2 - \frac{(\q\vec{k})^2}{\vec{k}^2}-
\frac{(\q\p)^2}{\p\,^2} +\frac{(\q\p)(\p
\vec{k})(\q\vec{k})}{\p\,^2 \vec{k}^2} \right] .  \nonumber
\ea
This expression can be written in terms of integrals already
solved in the course of the one loop calculation, and the
outcome is
\begar
 \Pi_{4.b} &=& -\frac{i g^4 \q\,^2 N^2 \d_{ab}}{24 \pi^4} [
\CD^2 + 2 \CD (\frac{7}{3}-2\ln 2) -2 \CD \ln \q\,^2
\plabel{Pi3b}\\ & & +2(\g-\ln 4\pi - \frac{14}{3} +4 \ln 2)
\ln\q\,^2 + 2 \ln^2 \q\,^2 + const+O(\e)]. \nonumber
\ea
Eq. \pref{Pi3b} contains overlapping divergences and two graphs
like 3.a with one or the other vertex replaced by a counterterm
have to be added. After that, only an additive infinity survives
which is subtracted by our usual renormalization prescription.
Graph 4.b has a net contribution which is proportional to
$\ln^2$:
\beg \plabel{Pserg}
 \Pi_{4.b}^{ren} = -i\frac{N^2 g^4 \q\,^2}{24\pi^4}\ln^2
\frac{\q\,^2}{\vec{\m}^2}.
\ee
The last two loop correction we are considering is the graph in
fig. 4.c, whose contribution to the Coulomb gluon propagator can
be written as
\newpage
\begar
 \Pi_{4.c} &=& -6ig^4 N^2 \int \frac{d^Dr}{(2\pi)^D} \int
\frac{d^Dk}{(2\pi)^D} (q-k)^r q^l \frac{1}{(\q-\vec{k})^2
(\q-\vec{k}-\vec{r})^2 r^2 k^2} \times \nonumber \\ & &  \qquad
\times \left( \d^{rl}-\frac{k^rk^l}{\vec{k}^2}
-\frac{r^rr^l}{\vec{r}\,^2}+\frac{r^r(\vec{r}\vec{k})k^l}{
\vec{r}\,^2\vec{k}^2}
\right).  \plabel{Pi3c}
\ea
Eq. \pref{Pi3c} can be evaluated entirely in terms of gamma
functions by a somewhat lengthy calculation, but following the
same steps as above.  The analytic result is
\begar
 \Pi_{4.c} &=& -ig^4 \frac{3N^2}{2(8\pi^2)^2} (\q\,^2)^{1-2\e}
(4\pi)^{2\e} \frac{\G{(\e)} \G{(2\e)} \G^2(2-\e)
\G(\frac{1}{2}-\e) \G(\frac{1}{2}-2\e)} {\G(\frac{5}{2}-2\e)
\G(\frac{5}{2}-3\e) \G(1+\e)} (1-2\e)(1-4\e).\nonumber \\
\ea
Expanding in terms of $\e$ and properly renormalizing the result
we finally obtain
\beg
 \Pi_{4.c} = \frac{i g^4 N^2}{48 \pi^4} (-\frac{4}{3}
\ln\frac{\q\,^2}{\vec{\m}^2}+\ln^2\frac{\q\,^2}{\vec{\m}^2})
\ee
In table 3 we collect the results for the self energy parts of
fig. 4.a to 4.c, apart from a factor $-i\frac{N^2 g^4}{24 \pi^4}
\q\,^ 2 $\\
\centerline{\begin{tabular}{c|c}
graph & $\Pi $ \\ \hline 4.a & $ \frac{3}{4}  \ln
\frac{\q\,^2}{\vec{\m}^2}$\\ 4.b & $ \ln^2
\frac{\q\,^2}{\vec{\m}^2} $ \\ 4.c & $
\frac{2}{3}\ln\frac{\q\,^2}{\vec{\m}^2} - \frac{1}{2}
\ln^2\frac{\q\,^2}{\vec{\m}^2} $
\end{tabular}}\\[.1cm]
\centerline{Tab. 3}\\[.3cm]
The full calculation of the gluon self energy to two loops seems
to be very involved in the Coulomb gauge. This is the more so,
if massive fermions are included. However, in view of the
results from the one loop calculation with massive fermions we
may expect that nonvanishing masses tend to decrease the
importance of such terms in practice to something that would be
de facto numerically equivalent to $O(\a^5)$.

We try to circumvent these problems, for the time being, by the
following argument, which also includes the three 'massless'
quarks u,d,s.  Because of the Ward identity for the
Coulomb-vertex, it is clear from the theory of renormalization
group that the same corrections can be obtained by expanding the
running coupling constant with a two loop (gluons+u,d,s) input
for the latter which provides also 'nonleading' logarithmic
contributions.  Our present calculation in any case illustrates
on the one hand the procedure to be followed in a systematic BS
perturbation theory. On the other hand, we believe that
especially the computation methods for the notoriously difficult
Coulomb gauge may be useful elsewhere.

The beta function to two loops is renormalization scheme
independent for massless quarks \pcite{Muta} and its two loop
part has been calculated some time ago \pcite{Jon}:
\begar
 \b(g) &=& -\b_0 g^3 - \b_1 g^5 - ...\\ \b_0 &=& \frac{1}{(4
\pi)^2} (11-\frac{2}{3} n_f)\\ \b_1 &=& \frac{1}{(4 \pi)^4}
(102-\frac{38}{3} n_f).
\ea
Here $n_f$ is the number of effective (massless) flavours and
$\b(g)$ is the solution of
\begar
 \ln \frac{\sqrt{-q^2}}{\m} &=& \int_g^{\bar{g}}
\frac{dg4}{\b(g)},
\ea
which reads up to two loops
\beg
  \ln \frac{\sqrt{-q^2}}{\m} = \frac{1}{2 \b_0} \left[
\frac{1}{\bar{g}^2}-\frac{1}{g^2} + \frac{\b_1}{\b_0} \ln
\frac{\bar{g}^2 (\b_0+\b_1 g^2)}{g^2 (\b_0+\b_1 \bar{g}^ 2)}
\right].
\ee
Considering this as an equation for $\bar{g}=g(\q\,^2)$  we
'undo' the renormalization group improvement by expanding with
'small' $g^2 \propto \a$ (cf. eq. \pref{alpha} ):
\begar
 \a(\q\,^2)&=& \a \left\{ 1- \a \frac{33-2 n_f}{16 \pi}\ln
\frac{\q\,^2}{\m^2}+ \right.\\ & &  \qquad + \left.
\frac{\a^2}{(16 \pi)^2}[(33-2 n_f)^2 \ln^2 \frac{\q\,^2}{\m^2}-
9(102-\frac{38}{3}n_f)\ln \frac{\q\,^2}{\m^2}] \right\}
\nonumber
\ea
Clearly the one loop term agrees with the calculation in
subsect.  4.1 in the limit $m_j \to 0$. Nevertheless, this term
should not be considered within the present argument because its
error has the same order of magnitude as contributions of
$O(\a^4)$ .  For the computation of the rest we need the
expectation value of $\frac{\ln^2 \frac{\q\,^2}{\m^2}}{\q\,^2}$.
This integral can be done analytically (Appendix A) and the
result is:
\begar
\<\frac{ \ln^2 \frac{\q\,^2}{\m^2}}{\q\,^2} \> &=& \frac{m \a}
{2 \pi n^2}  \{ \frac{\pi^2}{12} +
\Psi_2(n+l+1)+s_{nl}+[\Psi_1(n+l+1)+\g+ \ln \frac{\m n}{\a m}]^2
\}  \plabel{eln2}
\ea
with
\begar
s_{nl} &=& \frac{2 (n-l-1)!}{(n+l)!} \sum_{k=0}^{n-l-2}
\frac{(2l+1+k)!} {k! (n-l-1-k)^2}. \nonumber
\ea
With eq. \pref{eln2} we obtain for the mass shift, induced by
the leading logs of the two loop vacuum polarization of the
Coulomb gluon a contribution:
\begar
 \D M_{2loop} &=& -\frac{m \a^4}{128 \pi^2 n^2} \left\{ 27^2[
\frac{\pi^2}{12} + \Psi_2(n+l+1)+s_{nl}+(\Psi_1(n+l+1)+\g+ \ln
\frac{\m n}{\a m})^2] \right. + \nonumber \\ & & \qquad \qquad
\qquad+ \left.288 (\Psi_1(n+l+1) + \g +\ln \frac{\m n}{\a m})
\right\}.  \plabel{dM2l}
\ea
In this expression we have set the number of effective flavours
equal to 3 as dictated by the number of sufficiently light
quarks.  Whether eq. \pref{dM2l} really represents the full two
loop quark- gluon vacuum polarization, numerically consistent
with other terms $O(\a^4)$, must still be checked in a
calculation of the Coulomb gluon's self-energy to two loop order
in the Coulomb gauge, i.e. going beyond the sample calculation
here.

\subsection{QCD 2-loop Box Graphs}

It would be incorrect to extrapolate from the QED case the
absence of corrections of $O(\a^4)$, because gluon splitting
allows new types of graphs.  Our first example of a QCD box
graph is fig. 5.a. Between the nonrelativistic projectors
$\l^{\pm}$ of the wave functions the perturbation kernel from
this graph can be written effectively as
\begar
 -i H_{5.a} &=&-12 i g^6 \int \frac{d^4t}{(2\pi)^4}
\frac{d^4k}{(2\pi)^4}
\frac{(k_0+p_1^0+m)(q^0-t^0+p_2^0-m)}{[(p_1-k)^2-m^2]
[(p_2-t)^2-m^2]
\vec{k}\,^2 (\q-\vec{k})^2 \vec{t}\,^2 (\q-\vec{t})^2} \times
\nonumber \\ & & \quad \times \frac{1}{(t-k)^2}
\left(-(\q-\vec{k})\vec{k}+
\frac{[(\vec{t}-\vec{k})(\q-\vec{k})][\vec{k}
(\vec{t}-\vec{k})]} {(\vec{t}-\vec{k})^2} \right).
\plabel{Hgraph}
\ea
After performing the integrations over the zero components $t^0$
and $k^0$ the external momenta can be set to $(m,\vec{0})$. This
is justified a posteriori by the finiteness of the remaining
terms.  The resulting expression will thus only depend on $\q$
and $m$. The leading contribution seems to be
\beg
 H_{5.a} \propto \frac{\a^3}{ |\q|^2}, \plabel{H5a}
\ee
but a really reliable estimate or even a calculation of the
coefficient of the leading term is not available yet.
Supplementing the usual three powers of $\a$ from the wave
functions for the computation of energy levels, we see that the
graph 5.a indeed gives a contribution $O(\a^4)$. The qualitative
result \pref{H5a} had been noted already in
\pcite{Dun}, \pcite{Fisch} and \pcite{Fein}.
It should be noted, however, that also e.g. the graph 5.b may
yield a contribution of the same structure.  A similar graph
with crossed Coulomb lines (fig. 5.c) does not contribute
because of the fact that the group theoretic factor vanishes.

Now we investigate the 'X` graph in fig. 5.d for possible new
contributions.  As in the calculation of fig. 5.a it can be
simplified  to give
\begar
 -i H_{5.d} &=& 3 i g^6 \int \frac{d^4t}{(2\pi)^4}
\frac{d^4k}{(2\pi)^4}
\frac{(k_0+p_1^0+m)(t^0+p_2^0-m)}{[(k+p_1)^2-m^2][(t+p_2)^2-m^2]
\vec{k}\,^2 (\q-\vec{k})^2} \times  \nonumber \\ & & \quad
\times \frac{1}{t^2 (q-t)^2} \left( 1+
\frac{[\vec{t}(\q-\vec{t})]^2} {\vec{t}\,^2 (\q-\vec{t})^2}
\right).  \plabel{Xgraph}
\ea
The integration over $k$ yields a $1/|\q|$ divergence if $\q$
and $\p$ tend to zero. Contributions within the order of
interest can therefore only come from possible poles after the
$t$-integration. For simplicity we consider the part of eq.
\pref{Xgraph} from the factor one in the second line:
\begar
 I_t &\equiv& \int \frac{d^4t}{(2\pi)^4}
\frac{t^0+p_2^0+m}{[(t+p_2)^2-m^2] t^2 (q-t)^2} = \nonumber\\
&=& \frac{-i}{(4\pi)^2} \int_0^1dx \int_0^{1-x} dy \frac{y q^0
-x p_2^0+p^0+m} {(y q-x p)^2-y q^2 -x(p^2-m^2)} \approx
\nonumber\\ &\approx&  \frac{im}{(4\pi)^2} \int_0^1dx
\int_0^{1-x} dy \frac{x} {x^2 m^2+ y(1-x-y)\q\,^2} + O(\a)=
\nonumber\\ &=& \frac{-i}{(4\pi)^2m} \ln \frac{|\q|}{m} + O(\a).
\ea
Thus the part of graph 5.d, specified above, has a leading
contribution of $$ H_{5.d,1} = \frac{g^6 N}{32(4\pi)^2 |\q| m}
\ln \frac{|\q|}{m} $$ as $\q \to 0$ and therefore contributes to
$O(\a^5 \ln \a)$. The second part of graph 5.c gives a similar
contribution, only with a different numerical factor.  We
conclude that - in contrast to the QED case \pcite{Fult}- box
graphs may contribute to $O(\a^4)$. As illustrated by the
explicit calculations above, to $O(\a^5)$ beside abelian QED
type corrections \cite{Male,Fult}, a host of further non-abelian
contributions can be foreseen.

\subsection{QED Correction}

As a rule, one is not forced to consider also electromagnetic
effects in QCD calculations, but at high energies the strong
coupling decreases, and in the case of toponium we expect
$\a_s^2$ to be of the same order as $\a_{QED}$.  We can obtain
this contribution by solving the BR equation for the sum of an
QED and an QCD Coulomb gluon which results in the energy levels
\begar
 P_0 =M_n^0 &=& 2m \sqrt{1-\frac{(\a+\a_{QED} Q^2)^2}{4n^2}}
\approx  \\ &\approx& 2m - m \frac{\a^2}{4n^2} - \frac{m \a
\a_{QED} Q^2}{2 n^2} - m \frac{\a^4}{64n^4}  -\frac{m \a_{QED}^2
Q^4}{4 n^2} + O(\a^6), \nonumber
\ea
where $Q$ is the electric charge of the heavy quark, that means
$2/3$ for toponium. Clearly even the 'leading' third term can
only be separated from the effect of the second one to the
extent that $\a(\m)$ and $\a_{QED}(\m)$ can be studied
separately with sufficient precision.

\subsection{Weak Corrections}

While also weak interactions can usually be neglected in QCD
calculations, this is not true in the high energy region. This
is because the weak coupling scales like $\sqrt{G_F m^2}$ and
this becomes comparable to the strong coupling if the fermion
mass $m$ is large. Even bound states through Higgs exchange may
be conceivable \pcite{Jain}. Therefore we have to consider weak
corrections and especially the exchange of a single Higgs or Z
particle, assuming for simplicity the standard model with
minimal Higgs sector.

The Higgs boson gives rise to the kernel
\beg
 -i H_{Higgs} = -i \sqrt{2} G_F m^2 \frac{1}{q^2-m_H^2}\approx i
\sqrt{2} G_F m^2 \frac{1}{\q\,^2+m_H^2}, \plabel{HH}
\ee
where the notation should be obvious.

Since we do not know the ratio $\a m/m_H$ which would allow some
approximations if that ratio is small, we calculate explicitly
the level shifts by transforming into coordinate space. As in
Appendix A, we express the Laguerre polynomials in terms of
differentiations of the generating function, do the integration
and perform the differentiation afterwards to obtain an
expression valid for arbitrary levels and Higgs boson masses:
\beg
\D M_{Higgs} = -m \frac{G_F m^2 \a}{4 \sqrt{2} \pi} I_{nl}(
\frac{\a m}{n m_H}) \plabel{dMH}
\ee
with
\begar
 I_{nl}(a_n) &\equiv& \frac{a_n^{2l+2}}{n^2 (1+a_n)^{2n}}
\sum_{k=0}^{n-l-1} {n+l+k \choose k} {n-l-1 \choose k}
(a_n^2-1)^{n-l-1-k}
\ea
As an illustration some explicit results for the lowest levels
are given in tab.4.
\renewcommand{\arraystretch}{1.4}\\[.5cm]
\centerline{\begin{tabular}{c|cccccc}
 $n$ & 1 & 2 & 2 & 3 & 3 & 3 \\ \hline $l$ & 0 & 0 & 1 & 0 & 1 &
2 \\ \hline $I_{nl}(a_n)$ & $\frac{a_1^2}{(1+a_1)^2}$ &
$\frac{a_2^2(2+a_2^2)}{4(1+a_2)^4}$ & $\frac{a_2^4}{4(1+a_2)^4}$
& $\frac{a_3^2(3+6a_3^2+a_3^4)}{9(1+a_3)^6}$ &
$\frac{a_3^4(4+a_3^2)}{9(1+a_3)^6}$ & $\frac{a_3^6}{9(1+a_3)^6}$
\end{tabular}}

\vspace*{0.3cm}
\centerline{Tab. 4}
\vspace{.5cm}
It is evident that Eq. \pref{dMH} will give a contribution of
order $G_F m^2 \a^3$ if the Higgs mass is comparable to the mass
of the heavy quark and should therefore be taken into account in
a consistent treatment of heavy quarkonia spectra to numerical
order $O(\a^4)$.

Next we consider the contributions of the neutral current, the
single Z-exchange and Z-annihilation.

For the Z-exchange we obtain
\begar
 H_Z &=& -4 \sqrt{2} G_F m_Z^2 [\g^{\m}(g_1+g_2 \g_5)]_{\s\s'}
\frac{g_{\m\n}-\frac{q_{\m}q_{\n}}
{m_Z^2}}{q^2-m_Z^2}[\g^{\n}(g_1+g_2 \g_5)]_{\r'\r} \plabel{HZ}
\ea
with
\begar
 g_1 &=& \frac{1}{2} T^f_3-Q_f \sin^2 \Theta_w, \\ g_2 &=&
-\frac{1}{2} T^f_3,
\ea
where $T_3^f$ is the eigenvalue of the diagonal SU(2) generator
for the fermion $f$. If $f$ is the top quark then $T^f_3=1$.

Because $m_Z$ is expected to be larger than   $\a m$, it is
possible to write the leading term of the mass shift resulting
from
\pref{HZ} in the form
\begar
 \D M_Z &=& -4 \sqrt{2} G_F m_Z^2 \< \frac{g_2^2(3-2
\vec{S}\,^2) -g_1^2} {q\,^2-m_Z^2} \> = \\ &=& m \frac{G_F m^2
\a^3}{\sqrt{2} \pi n^3} (g_2^2(3-2 \vec{S}\,^2) -g_1^2) \d_{l0}
\nonumber
\ea
where $\vec{S}$ is total spin of the quark-antiquark system.
Therefore this expression gives rise to a singlet triplet
splitting within the order of interest.

In contrast to the gluon annihilation there is no colour
structure involved in the Z-annihilation graph. Therfore this
process does contribute. The corresponding energy shift is
easily evaluated:
\begar
 \D M_{S=0} &=& \frac{6 G_F m^2 g_2^2}{\sqrt{2} \pi} \frac{m
\a^3} {n^3} \d_{l0} \\ \D M_{S=1} &=&
-\frac{g_1^2}{g_2^2}\frac{m_Z^2}{4m^2-m_Z^2} \D M_{S=0}
\ea
Therefore this contribution also yields a singlet triplet
splitting.

\subsection{Schwinger-Christ-Lee Terms}

As mentioned in section \pref{Form}, nonlocal interactions have
to be added to the Lagrangian of QCD in Coulomb gauge.  We are
not aware of any previous attempt to look whether these terms
give contributions to bound state problems. In fact, by explicit
calculation we found though that the Schwinger-Christ-Lee terms
$v_1$ and $v_2$ in \pref{Lag} only give rise to corrections to
the propagator of the {\it transverse} gluon, and therefore can
actually be neglected to $O(\a^4)$.

By analogy to the second ref. \pcite{Schw} we calculate the
$v_1$ term for SU(N) to $O(g^4)$
\begar
 v_1 &=& -g^4 \frac{N^2}{16} \int d^3r d^3r' d^3r''
A_i^c(\vec{r}\,') K_{ij}(\vec{r}-\vec{r}\,')
K_{jk}(\vec{r}-\vec{r}\,'') A_k^c(\vec{r}\,'')\\ K_{ij}(\r) &=&
\frac{1}{4 \pi |\r|} \left[ \frac{\d_{ij}}{3}
\d(\vec{\r})-\frac{1}{4 \pi |\vec{\r}|^5} (3 \r_i \r_j -
\vec{\r}\,^2 \d_{ij}) \right]. \nonumber
\ea
This corrects the gluon propagator by
\beg
 \d G_{\m\n}^{ab}(x_1,x_2) = -\frac{1}{Z[0]} \frac{\d^2}{\d
J_{\m}^a(x_1) \d J_{\n}^b(x_2)} \frac{i g^4 N^2}{16} \int d^4x
\int d^3r d^3r' d^3r'' \frac{\d}{\d J_i^c} K_{ij} K_{jk}
\frac{\d}{\d J_k^c} Z_0[J].
\ee
In momentum space $\d G$ can be calculated by using dimensional
regularization to give
\beg
 \d G_{mn}^{ab} (q,q') = (2\pi)^4 i \d^{ab} \d(q-q') \frac{ g^4
N^2}{8^5} \frac{\q^2}{q^2} \frac{1}{q^2}(-\d_{mn}+\frac{q_m
q_n}{\q^2}),
\ee
which means that we have a mass shift with the same structure as
the one transverse gluon exchange (cf. sect. \pref{tree}), only
suppressed by two more orders in $\a$.

Since the second term $v_2$ also represents a correction to the
propagator of the transverse gluon, it can be estimated by the
same method to contribute only in higher orders of $\a$ as well.

\begin{appendix}

\setcounter{equation}{0}

\section{Calculation of expectation values}

In sect. 4 and 6 we needed the expectation values of logarithmic
potentials between Schr\"odinger wave functions. They can be
obtained by
\beg
 \< \frac{\ln^nr}{r} \> = \frac{d^n}{d\l^n} \< r^{\l-1} \>
\big|_{\l=0},
\ee
if the expectation value $\< r^{\l-1} \> $ is known
analytically.  This expression can be calculated by using the
representation
\beg
 L_{n-l-1}^{2l+1}(\r) = \lim_{z\to 0} \frac{1}{(n-l-1)!}
\frac{d^{n-l-1}}{dz^{n-l-1}} (1-z)^{-2l-2} e^{\r \frac{z}{z-1}}
\ee
of the Laguerre polynomials. This allows an easy evaluation of
the integrations and the remaining differentiations can be done
with some care:
\beg
 \< r^{\l-1} \> = \frac{(\a m)^{1-\l }}{2 n^{2-\l } }
\frac{(n-l-1)!}{(n+l)!} \G(2l+2+\l) \sum_{k=0}^{n-l-1} {\l
\choose   n-l-1-k }^2 {-2l-2-\l \choose k } (-1)^k
\ee
Using now the Fourier transformations \pcite{Gelf}
\begar
 F[ \frac{\ln\frac{\q\,^2}{\m^2}}{\q\,^2}] &=& -\frac{\g+\ln \m
r}{2\pi r}\\ F[ \frac{\ln^2\frac{\q\,^2}{\m^2}}{\q\,^2}] &=&
\frac{1}{2\pi r} [\frac{\pi^2}{6} + 2(\g+\ln \m r)^2]
\ea
one arrives immediately at eq.\pref{dMg} and \pref{eln2},
respectively.

\end{appendix}

\newpage

\u{Figure Captions}
\begin{itemize}
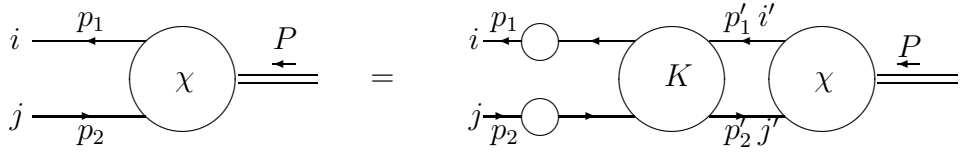
\item [Fig.1:] BS-equation for bound states
\item [Fig.2:] Tree graphs (broken lines represent Coulomb
gluons, curly lines depict transverse
       gluons, wavy lines represent a general gluon and solid
lines stand for fermions).
\item [Fig.3:] One loop graphs and vertex corrections
\item [Fig.4:] Two loop vacuum polarization
\item [Fig.5:] Two loop box graphs
\end{itemize}
\newpage

\unitlength1.0cm
\begin{picture}(18,5)
  \put(.5,3.5){\line(1,0){1.5}}
\put(3.24,3.05){\line(1,0){1.05}} \put(.5,2.5){\line(1,0){1.5}}
\put(2.5,3){\circle{1.5}}\put(3.24,2.95){\line(1,0){1.05}}
\put(1.5,3.5){\vector(-1,0){0.3}}    \put(2.4,2.9){$\chi$}
\put(1,2.5){\vector(1,0){0.3}} \put(1.1,3.7){$p_1$}
\put(3.7,3.4){$P$} \put(1.1,2.2){$p_2$}
\put(4,3.2){\vector(-1,0){0.3}} \put(0.2,3.4){$i$}
\put(0.2,2.4){$j$}           \put(5,2.9){=}

  \put(6.5,3.5){\line(1,0){.5}} \put(7.25,3.5){\circle{.5}}
\put(7.5,3.5){\line(1,0){1}} \put(6.5,2.5){\line(1,0){.5}}
\put(7.25,2.5){\circle{.5}} \put(7.5,2.5){\line(1,0){1}}
\put(7,3.5){\vector(-1,0){0.3}}
\put(8.2,3.5){\vector(-1,0){0.3}}
\put(6.5,2.5){\vector(1,0){0.3}}
\put(7.7,2.5){\vector(1,0){0.3}} \put(9.5,3.5){\line(1,0){1}}
\put(9,3){\circle{1.5}} \put(9.5,2.5){\line(1,0){1}}
\put(8.9,2.9){$K$} \put(11.74,3.05){\line(1,0){1.05}}
\put(11,3){\circle{1.5}} \put(11.74,2.95){\line(1,0){1.05}}
\put(10.9,2.9){$\chi$} \put(10.2,3.5){\vector(-1,0){0.3}}
\put(12.3,3.2){\vector(-1,0){0.3}}
\put(9.7,2.5){\vector(1,0){0.3}}
\put(12,3.3){$P$} \put(6.6,3.7){$p_1$}
\put(9.7,3.7){$p_1'$} \put(6.6,2.2){$p_2$}
\put(9.7,2.2){$p_2'$} \put(6.3,3.4){$i$}
\put(10.15,3.7){$i'$} \put(6.3,2.4){$j$}
\put(10.15,2.2){$j'$} \put(5,1){Fig. 1} \end{picture}
\unitlength1cm
\centerline{\begin{picture}(9,3)
\put(1,2){\line(1,0){3}} \multiput(2,2)(1.5,0){2}{
\vector(-1,0){.3}}
\put(1,.5){\line(1,0){3}} \multiput(1.7,.5)(1.5,0){2}{
\vector(1,0){.3}}
\put(1,.7){$\r$} \put(4,.7){$\r'$}
\put(1,2.2){$\s$} \put(4,2.2){$\s'$}
\multiput(2.5,.65)(0,.3){5}{\line(0,-1){.2}}
\put(2.5,.5){\circle*{.1}} \put(2.5,2){\circle*{.1}}
\put(4.5,1.23){$-$}
\put(5,2){\line(1,0){3}} \multiput(6,2)(1.5,0){2}{
\vector(-1,0){.3}}
\put(5,.5){\line(1,0){3}} \multiput(5.7,.5)(1.5,0){2}{
\vector(1,0){.3}}
\put(4.8,.7){$\r$} \put(8.1,.7){$\r'$} \put(8.5,.7){;}
\put(4.8,2.2){$\s$} \put(8.1,2.2){$\s'$}
\multiput(6.5,.65)(0,.3){5}{\circle*{.15}}
\put(6.6,1.2){$K_{BR}$}
\put(6.5,.5){\circle*{.1}} \put(6.5,2){\circle*{.1}}
\put(4.2,0){2.a}
\end{picture}}
\input feynman
\newcommand{\ds}[4]{\global\seglength=1416 \global
 \gaplength=850 \drawline\scalar[#1\REG](#2,#3)[#4] }
\centerline{\begin{picture}(12000,8000)
\drawline\fermion[\E\REG](1000,7000)[10000]
\multiput(2200,7000)(7000,0){2}{\vector(-1,0){.3}}
\drawline\fermion[\E\REG](1000,1000)[10000]
\multiput(2700,1000)(7000,0){2}{\vector(1,0){.3}}
\drawline\gluon[\N\FLIPPEDCURLY](6000,1000)[7]
 \put(6000,7000){\circle*{250}}
\put(6000,1000){\circle*{250}}
\put(5000,-500){2.b}
\end{picture}
\begin{picture}(12000,8000)
\drawline\photon[\E\REG](3000,4000)[6]
\drawline\fermion[\NW\REG](\photonfrontx,\photonfronty)[2500]
\put(\photonfrontx,\photonfronty){\circle*{250}}
\put(\photonfrontx,\photonfronty){\vector(-1,1){1000}}
\drawline\fermion[\SW\REG](\photonfrontx,\photonfronty)[2500]
\put(\photonfrontx,\photonfronty){\vector(1,1){-800}}
\drawline\fermion[\NE\REG](\photonbackx,\photonbacky)[2500]
\put(\photonbackx,\photonbacky){\circle*{250}}
\put(\photonbackx,\photonbacky){\vector(-1,-1){-800}}
\drawline\fermion[\SE\REG](\photonbackx,\photonbacky)[2500]
\put(\photonbackx,\photonbacky){\vector(1,-1){1000}}
\put(5500,-500){2.c}
\end{picture}}
\centerline{Fig. 2}

\newpage

\begin{picture}(20000,18000)
\drawline\fermion[\E\REG](2000,15500)[16000]
\drawline\fermion[\E\REG](2000,2000)[16000]
\multiput(3500,15500)(12000,0){2}{\vector(-1,0){.3}}
\multiput(4000,2000)(12000,0){2}{\vector(1,0){.3}}
\ds{\N}{10000}{2000}{6}
\put(10000,15500){\circle*{250}}
\put(10000,2000){\circle*{250}}
\drawloop\gluon[\E 5](10000,11000)
\put(10000,11000){\circle*{250}}
\put(10000,\gluonbacky){\circle*{250}}
\put(9500,0){3.a}
\end{picture}
\begin{picture}(20000,18000)
\drawline\fermion[\E\REG](2000,15500)[16000]
\drawline\fermion[\E\REG](2000,2000)[16000]
\multiput(3500,15500)(12000,0){2}{\vector(-1,0){.3}}
\multiput(4000,2000)(12000,0){2}{\vector(1,0){.3}}
\ds{\S}{10000}{15500}{2}
\ds{\N}{10000}{2000}{2}
\put(10000,15500){\circle*{250}}
\put(10000,2000){\circle*{250}}
\drawloop\gluon[\E 5](10250,11000)
\drawloop\gluon[\W 5](9750,6125)
\put(9750,6125){\line(1,0){500}}
\put(9750,11000){\line(1,0){500}}
\put(10000,11000){\circle*{250}}
\put(10000,6125){\circle*{250}}
\put(9500,0){3.b}
\end{picture}

\begin{picture}(20000,18000)
\drawline\fermion[\E\REG](2000,15500)[16000]
\drawline\fermion[\E\REG](2000,2000)[16000]
\multiput(3500,15500)(12000,0){2}{\vector(-1,0){.3}}
\multiput(4000,2000)(12000,0){2}{\vector(1,0){.3}}
\ds{\S}{10000}{15500}{2}
\ds{\N}{10000}{2000}{2}
\put(10000,15500){\circle*{250}}
\put(10000,2000){\circle*{250}}
\put(10000,8750){\circle{6000}}
\put(8000,8850){\vector(0,1){.3}}
\put(12000,8650){\vector(0,-1){.3}}
\put(10000,10750){\circle*{250}}
\put(10000,6750){\circle*{250}}
\put(9500,0){3.c}
\end{picture}

\begin{picture}(20000,12000)
\drawline\fermion[\E\REG](2000,8000)[16000]
\drawline\fermion[\E\REG](2000,2000)[16000]
\multiput(3500,8000)(12000,0){2}{\vector(-1,0){.3}}
\multiput(4000,2000)(12000,0){2}{\vector(1,0){.3}}
\ds{\N}{10000}{2000}{3}
\put(10000,8000){\circle*{250}}
\put(10000,2000){\circle*{250}}
\drawloop\gluon[\N 5](7500,8000)
\put(7500,8000){\circle*{250}}
\put(\gluonbackx,\gluonbacky){\circle*{250}}
\put(19000,0){3.d}
\end{picture}
\begin{picture}(20000,12000)
\drawline\fermion[\E\REG](2000,8000)[16000]
\drawline\fermion[\E\REG](2000,2000)[16000]
\multiput(3500,8000)(12000,0){2}{\vector(-1,0){.3}}
\multiput(4000,2000)(12000,0){2}{\vector(1,0){.3}}
\ds{\N}{10000}{2000}{3}
\put(10000,8000){\circle*{250}}
\put(10000,2000){\circle*{250}}
\drawloop\gluon[\N 5](3000,8000)
\put(3000,8000){\circle*{250}}
\put(\gluonbackx,\gluonbacky){\circle*{250}}
\end{picture}

\begin{picture}(20000,18000)
\drawline\fermion[\E\REG](2000,2000)[16000]
 \multiput(4000,2000)(12000,0){2}{\vector(1,0){.3}}
\ds{\N}{10000}{2000}{3}
\put(\scalarbackx,\scalarbacky){\circle*{250}}
\drawline\gluon[\NE\REG](\scalarbackx,\scalarbacky)[6]
\drawline\scalar[\NW\REG](\scalarbackx,\scalarbacky)[4]
\drawline\fermion[\E\REG](2000,\gluonbacky)[16000]
\multiput(2500,\gluonbacky)(14500,0){2}{\vector(-1,0){.3}}
\put(\gluonbackx,\gluonbacky){\circle*{250}}
\put(3500,\gluonbacky){\circle*{250}}
\put(10000,2000){\circle*{250}}
\end{picture}
\begin{picture}(20000,18000)
\drawline\fermion[\E\REG](2000,2000)[16000]
\multiput(4000,2000)(12000,0){2}{\vector(1,0){.3}}
\ds{\N}{10000}{2000}{3}
\put(\scalarbackx,\scalarbacky){\circle*{250}}
\drawline\gluon[\NW\REG](\scalarbackx,\scalarbacky)[6]
\drawline\scalar[\NE\REG](\scalarbackx,\scalarbacky)[4]
\drawline\fermion[\E\REG](2000,\gluonbacky)[16000]
\multiput(2500,\gluonbacky)(14500,0){2}{\vector(-1,0){.3}}
\put(\gluonbackx,\gluonbacky){\circle*{250}}
\put(16500,\gluonbacky){\circle*{250}}
\put(10000,2000){\circle*{250}}
\put(-1500,0){3.e}
\end{picture}

 \begin{picture}(20000,18000)
\drawline\fermion[\E\REG](2000,2000)[16000]
 \multiput(4000,2000)(12000,0){2}{\vector(1,0){.3}}
\ds{\N}{10000}{2000}{3}
\put(\scalarbackx,\scalarbacky){\circle*{250}}
\drawline\gluon[\NW\REG](\scalarbackx,\scalarbacky)[6]
\put(\gluonbackx,\gluonbacky){\circle*{250}}
\drawline\gluon[\NE\REG](\scalarbackx,\scalarbacky)[6]
\drawline\fermion[\E\REG](2000,\gluonbacky)[16000]
\multiput(2500,\gluonbacky)(14500,0){2}{\vector(-1,0){.3}}
\put(\gluonbackx,\gluonbacky){\circle*{250}}
\put(10000,2000){\circle*{250}}
\put(9500,0){3.f}
\end{picture}
\begin{picture}(20000,18000)
\drawline\fermion[\E\REG](2000,2000)[16000]
\multiput(4000,2000)(12000,0){2}{\vector(1,0){.3}}
\drawline\photon[\NE\REG](6000,2000)[12]
\put(\photonbackx,\photonbacky){\circle*{250}}
\drawline\fermion[\E\REG](2000,\photonbacky)[16000]
\multiput(2500,\photonbacky)(14500,0){2}{\vector(-1,0){.3}}
\put(\photonbackx,2000){\circle*{250}}
\drawline\photon[\NW\REG](\photonbackx,2000)[12]
\put(\photonbackx,\photonbacky){\circle*{250}}
\put(6000,2000){\circle*{250}}
\put(9500,0){3.g}
\end{picture}

 \begin{picture}(20000,18000)
\drawline\fermion[\E\REG](2000,2000)[16000]
 \multiput(4000,2000)(12000,0){2}{\vector(1,0){.3}}
\ds{\N}{10000}{2000}{3}
\put(\scalarbackx,\scalarbacky){\circle*{250}}
\drawline\gluon[\NW\REG](\scalarbackx,\scalarbacky)[6]
\put(\gluonbackx,\gluonbacky){\circle*{250}}
\drawline\gluon[\NE\REG](\scalarbackx,\scalarbacky)[6]
\ds{\N}{\scalarbackx}{\scalarbacky}{3}
\drawline\fermion[\E\REG](2000,\gluonbacky)[16000]
\multiput(2500,\gluonbacky)(14500,0){2}{\vector(-1,0){.3}}
\put(\gluonbackx,\gluonbacky){\circle*{250}}
\put(10000,2000){\circle*{250}}
\put(10000,\fermionbacky){\circle*{250}}
\put(9500,0){3.h}
\end{picture}

\centerline{Fig. 3}

\newpage

\begin{picture}(20000,18000)
\drawline\fermion[\E\REG](2000,15500)[16000]
\drawline\fermion[\E\REG](2000,2000)[16000]
\multiput(3500,15500)(12000,0){2}{\vector(-1,0){.3}}
\multiput(4000,2000)(12000,0){2}{\vector(1,0){.3}}
\ds{\S}{10000}{15500}{6}
\put(10000,15500){\circle*{250}}
\put(10000,2000){\circle*{250}}
\drawloop\gluon[\E 5](10250,11000)
\drawloop\gluon[\W 5](9750,6125)
\put(9750,6125){\line(1,0){500}}
\put(9750,11000){\line(1,0){500}}
\put(10000,11000){\circle*{250}}
\put(10000,6125){\circle*{250}}
\put(9500,0){4.a}
\end{picture}
\begin{picture}(20000,18000)
\drawline\fermion[\E\REG](2000,15500)[16000]
\drawline\fermion[\E\REG](2000,2000)[16000]
\multiput(3500,15500)(12000,0){2}{\vector(-1,0){.3}}
\multiput(4000,2000)(12000,0){2}{\vector(1,0){.3}}
\ds{\S}{10000}{15500}{6}
\put(10000,15500){\circle*{250}}
\put(10000,2000){\circle*{250}}
\drawloop\gluon[\E 5](10000,8750)
\put(10000,\gluonbacky){\circle*{250}}
\drawloop\gluon[\W 5](10000,8750)
\put(10000,\gluonbacky){\circle*{250}}
\put(10000,8750){\circle*{250}}
\put(9500,0){4.b}
\end{picture}

\begin{picture}(20000,18000)
\drawline\fermion[\E\REG](2000,15500)[16000]
\drawline\fermion[\E\REG](2000,2000)[16000]
 \multiput(3500,15500)(12000,0){2}{\vector(-1,0){.3}}
\multiput(4000,2000)(12000,0){2}{\vector(1,0){.3}}
\ds{\S}{10000}{15500}{6}
\put(10000,15500){\circle*{250}}
\put(10000,2000){\circle*{250}}
\put(10000,9500){\line(1,0){250}}
\drawloop\gluon[\E 5](10250,9500)
\put(9750,\gluonbacky){\line(1,0){500}}
\put(10000,\gluonbacky){\circle*{250}}
\drawloop\gluon[\W 3](9750,\gluonbacky)
\drawline\gluon[\N \REG](\gluonbackx,\gluonbacky)[2]
\drawloop\gluon[\N 3](\gluonbackx,\gluonbacky)
\put(9750,\gluonbacky){\line(1,0){250}}
\put(10000,9500){\circle*{250}}
\put(10000,\gluonbacky){\circle*{250}}
\put(9500,0){4.c}
\end{picture}

\centerline{Fig. 4}

\newpage

\begin{picture}(20000,18000)
\drawline\fermion[\E\REG](2000,2000)[16000]
\multiput(3750,2000)(12750,0){2}{\vector(1,0){.3}}
\drawline\scalar[\N\REG](5000,2000)[6]
\drawline\fermion[\E\REG](2000,\scalarbacky)[16000]
\multiput(3750,\scalarbacky)(12750,0){2}{\vector(-1,0){.3}}
\put(\scalarbackx,\scalarbacky){\circle*{250}}
\put(5000,9000){\circle*{250}}
\drawline\gluon[\E\REG](5000,9000)[9]
\put(\gluonbackx,\gluonbacky){\circle*{250}}
\put(\gluonbackx,2000){\circle*{250}}
\drawline\scalar[\N\REG](\gluonbackx,2000)[6]
\put(\scalarbackx,\scalarbacky){\circle*{250}}
\put(5000,2000){\circle*{250}}
\put(9500,0){5.a}
\end{picture}
\begin{picture}(20000,18000)
\drawline\fermion[\E\REG](2000,2000)[16000]
\multiput(3750,2000)(12750,0){2}{\vector(1,0){.3}}
\drawline\gluon[\N\REG](5000,2000)[11]
\drawline\fermion[\E\REG](2000,\gluonbacky)[16000]
\multiput(3750,\gluonbacky)(12750,0){2}{\vector(-1,0){.3}}
\put(\gluonbackx,\gluonbacky){\circle*{250}}
\put(5000,8000){\circle*{250}}
\drawline\photon[\E\REG](5000,8000)[9]
\put(\photonbackx,\photonbacky){\circle*{250}}
\put(\photonbackx,2000){\circle*{250}}
\drawline\gluon[\N\REG](\photonbackx,2000)[11]
\put(\gluonbackx,\gluonbacky){\circle*{250}}
\put(5000,2000){\circle*{250}}
\put(9500,0){5.b}
\end{picture}

\begin{picture}(20000,18000)
\drawline\fermion[\E\REG](2000,2000)[16000]
\multiput(3750,2000)(12750,0){2}{\vector(1,0){.3}}
\drawline\photon[\N\REG](6000,2000)[6]
\newcounter{ax}
\newcounter{ay}
\setcounter{ax}{\photonbackx}
\setcounter{ay}{\photonbacky}
\drawline\photon[\E\REG](6000,\photonbacky)[8]
\newcounter{bx} \newcounter{by}
\setcounter{bx}{\photonbackx}
\setcounter{by}{\photonbacky}
\put(\photonfrontx,\photonfronty){\circle*{250}}
\drawline\photon[\N\REG](\arabic{bx},2000)[6]
\put(\photonbackx,\photonbacky){\circle*{250}}
\drawline\photon[\NW\REG](\arabic{bx},\arabic{by})[10]
\put(\photonbackx,\photonbacky){\circle*{250}}
\drawline\photon[\NE\REG](\arabic{ax},\arabic{ay})[10]
\drawline\fermion[\E\REG](2000,\photonbacky)[16000]
\multiput(3750,\photonbacky)(12750,0){2}{\vector(-1,0){.3}}
\put(\photonbackx,\photonbacky){\circle*{250}}
\put(\arabic{bx},2000){\circle*{250}}
\put(\photonbackx,\photonbacky){\circle*{250}}
\put(6000,2000){\circle*{250}}
\put(9500,0){5.c}
\end{picture}
\begin{picture}(20000,18000)
\drawline\fermion[\E\REG](2000,2000)[16000]
\multiput(3500,2000)(13500,0){2}{\vector(1,0){.3}}
\drawline\scalar[\NE\REG](4200,2000)[4]
\put(\scalarbackx,\scalarbacky){\circle*{250}}
\drawline\gluon[\NW\REG](\scalarbackx,\scalarbacky)[5]
\put(\gluonbackx,\gluonbacky){\circle*{250}}
\drawline\gluon[\NE\REG](\scalarbackx,\scalarbacky)[5]
\drawline\scalar[\SE\REG](\scalarbackx,\scalarbacky)[4]
\put(\scalarbackx,\scalarbacky){\circle*{250}}
\drawline\fermion[\E\REG](2000,\gluonbacky)[16000]
\multiput(3500,\gluonbacky)(13500,0){2}{\vector(-1,0){.3}}
\put(\gluonbackx,\gluonbacky){\circle*{250}}
\put(4200,2000){\circle*{250}}
\put(9500,0){5.d}
\end{picture}

\centerline{Fig. 5}

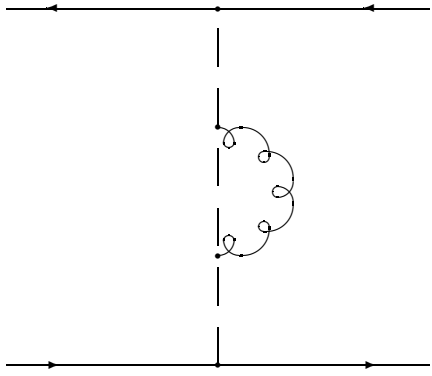
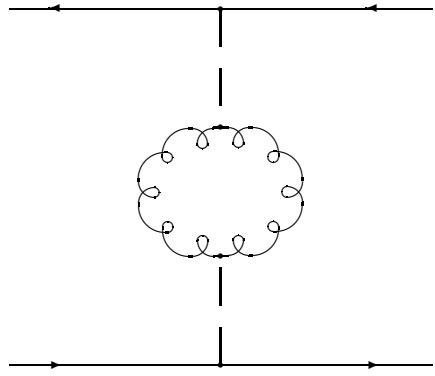
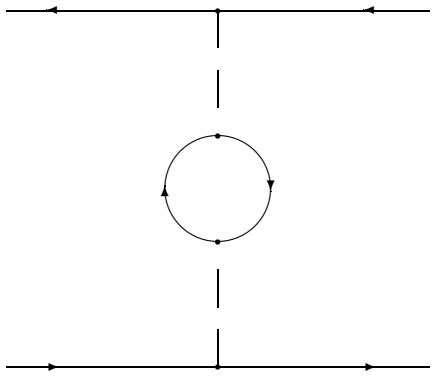
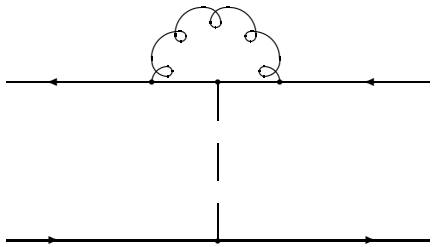
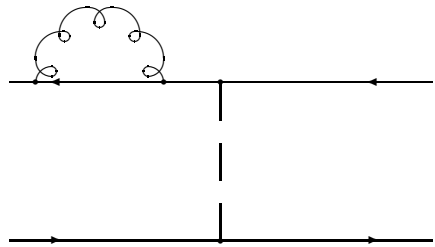
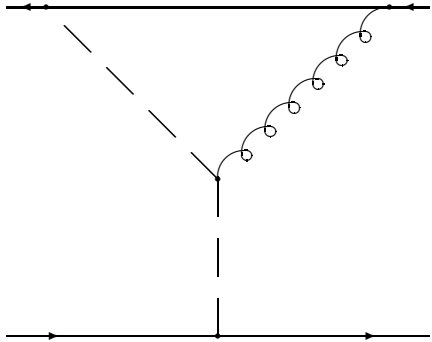
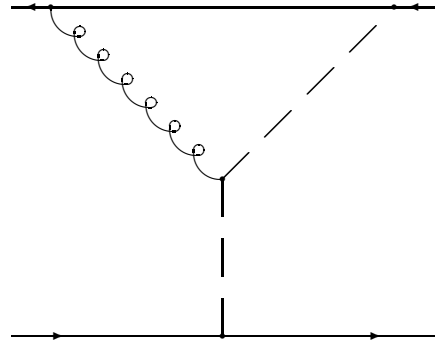
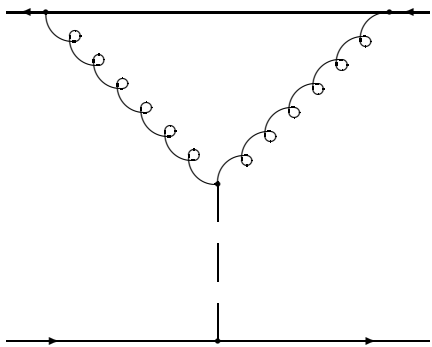
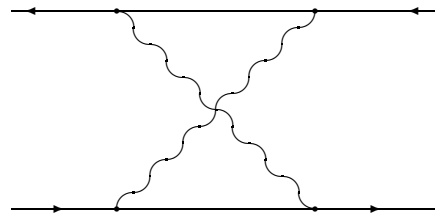
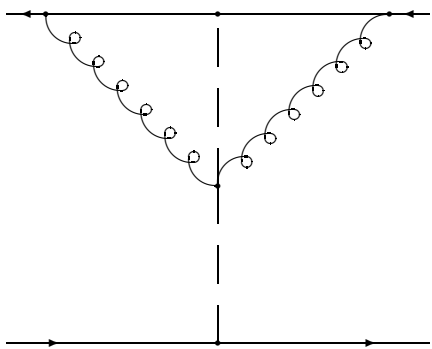
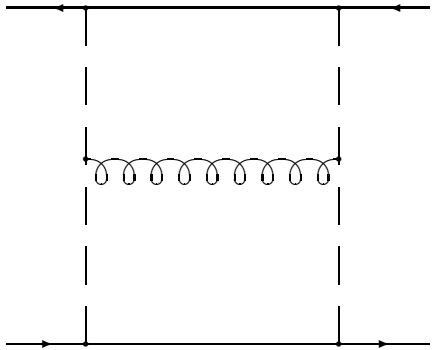
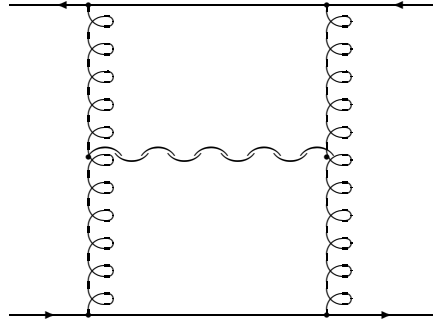
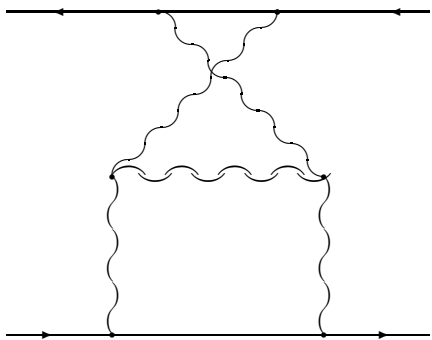
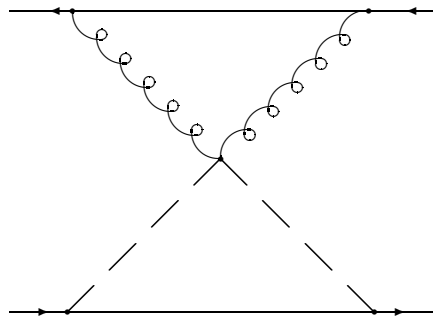
\end{document}